\documentclass[twocolumn,aps,showpacs,floatfix]{revtex4}

\usepackage{graphicx}% Include figure files
\usepackage{bm}% bold math
\begin{document}

\title{Production of a Fermi gas of atoms in an optical lattice.}

\author{G. Modugno, F. Ferlaino}
\altaffiliation[Permanent address: ]{5. Physikalisches Institut,
Universit\"{a}t Stuttgart, Pfaffenwaldring 57, 70550 Stuttgart,
Germany}
\author{R. Heidemann}
\author{G. Roati}
\author{M. Inguscio}
\affiliation{LENS and Dipartimento di Fisica, Universit\`a di Firenze, and INFM\\
 Via Nello Carrara 1, 50019 Sesto Fiorentino, Italy }

\date{\today}

\begin{abstract}
We prepare a degenerate Fermi gas of potassium atoms by
sympathetic cooling with rubidium atoms in a one-dimensional
optical lattice. In a tight lattice we observe a change of the
density of states of the system, which is a signature of quasi two
dimensional confinement. We also find that the dipolar
oscillations of the Fermi gas along the tight lattice are almost
completely suppressed.           
\end{abstract}


\maketitle

The combination of atomic Bose-Einstein condensates (BECs) and
periodic potentials created by light has recently allowed the
study of a variety of fundamental phenomena related to
phase-coherence and superfluidity \cite{kase,cata}, transport
\cite{morsch,philips}, and strongly correlated systems
\cite{bloch}. The corresponding interest in Fermi gases in optical
lattices is arising \cite{cirac,torma,albus}, especially in view
of possible application of lattices to the achievement and
detection of fermionic superfluidity.

In this work we study the production, and the basic static and
transport properties of a Fermi gas of atoms in a one-dimensional
(1D) optical lattice in the regime of tight confinement. We
prepare a sample of degenerate fermionic potassium atoms
($^{40}$K) in the lattice combined with a magnetic potential by
means of sympathetic cooling with bosonic rubidium ($^{87}$Rb). We
observe that the cooling maintains its efficiency also when the
atoms are confined in quasi 2D in the lattice sites. We detect the
reduced dimensionality of the Fermi gas under these conditions by
studying its momentum distribution. The comparison of the behavior
of the Fermi gas and the BEC in the lattice helps to confirm the
different transport properties expected for the two kinds of
quantum gases. In particular, we observe that the dipole
oscillations of the Fermi gas are strongly reduced in the tight
lattice.

The degenerate mixture is initially prepared in a magnetic
potential as described in Ref.\cite{fermibose}. The axial and
radial frequencies of the potential are $\omega_{a}=2\pi \times 24
(16.3)$~s$^{-1}$ and $\omega_{r}=2\pi\times 317 (215)$~s$^{-1}$
for K (Rb). Typically 5$\times$10$^{4}$ fermions are
sympathetically cooled to about 0.3$T_F$ ($T_F$=430~nK) in
coexistence with a BEC with a similar atom number.

\begin{figure}
\centerline{\includegraphics[width=8.5cm,clip=]{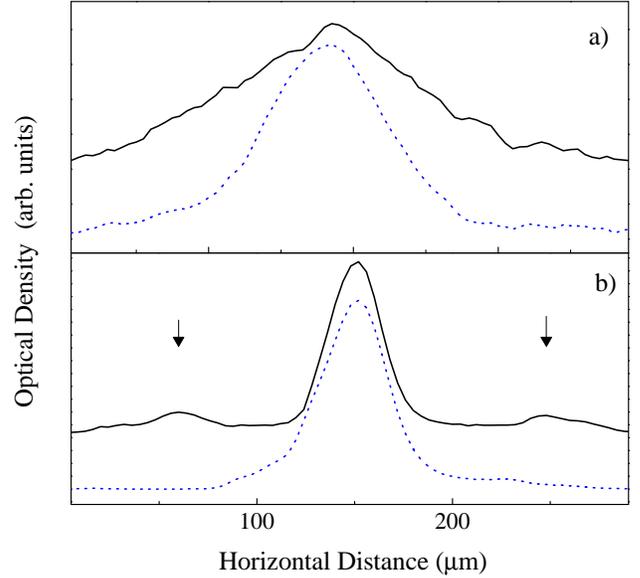}}
\caption{Axial profiles of the Fermi gas (a) and of the BEC (b)
after ballistic expansion from a 1D lattice combined with the
magnetic trap (continuous lines). The interference peaks in the
BEC profile (arrows) do not show up in the Fermi gas. The
expansion from the pure magnetic trap is shown for comparison
(dotted lines, the curves are offset for clarity). The expansion
time is 8 ms and 17 ms for the Fermi gas and the BEC,
respectively.} \label{fig:image}
\end{figure}

The lattice is created with a laser beam in a standing-wave
configuration, focused to a beam waist of 500~$\mu$m and aligned
along the weak axis of the magnetic potential. At a wavelength
$\lambda$=754~nm the lattice potential is repulsive and its depth
is about twice as large for K than for Rb. More precisely, the
depth for the two species is almost the same in units of the
recoil energy $U$=$s E_R$, with $E_R$=$\hbar^2 k^2/2 m$ and
$k$=$2\pi/\lambda$ \cite{noterecoil}.

We have investigated the loading and cooling of the Fermi gas in
the tightest lattice we could achieve, with $s$=8. We find that
the optimal loading is obtained by raising the lattice beam from
zero to full power in about 500~ms during the final stage of the
evaporation, in order to reach the full depth when the phase
transition for bosons occurs. By continuing the evaporation of
rubidium for approximately 1 s, we observe that the mixture cools
efficiently in the lattice to quantum degeneracy. The presence of
the lattice shows up in the characteristic interference peaks in
the momentum distribution of the BEC \cite{pedri}, as shown in
Fig.~\ref{fig:image}. Because of the broad momentum distribution,
this is not observed in fermions, which just show a faster axial
expansion because of the additional confinement provided by the
lattice.

\begin{figure}
\centerline{\includegraphics[width=8.5cm,clip=]{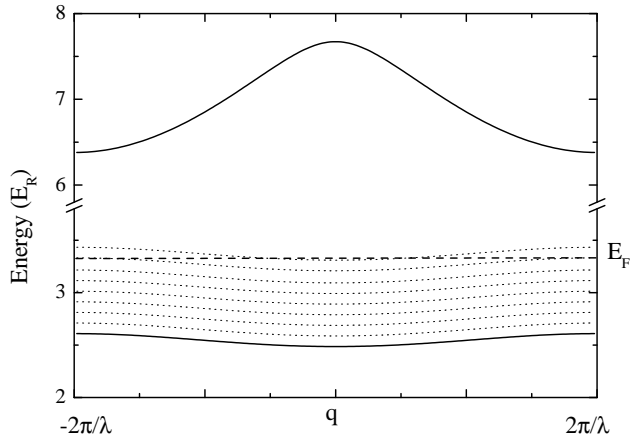}}
\caption{Calculated energy spectrum of the first two bands of the
1D lattice for $s$=8. Many bands, whose spacing has been
exaggerated for clarity, are actually present (dotted lines). The
dashed line represents the Fermi energy of the system.}
\label{fig:band}
\end{figure}

Previous experiments on bosons have shown that a tight lattice can
lead to a collection of BECs with a quasi-2D character, where the
reduced dimensionality affects the phase transition to quantum
degeneracy \cite{bec2d}. To understand the dimensionality of the
Fermi gas, we first need to study the energy spectrum of the
fermions in the combined potential. We initially use a simplified
model, in which we replace the axial harmonic confinement with a
boxlike potential. Within this approximation the atoms experience
a uniform lattice with $N_l$ sites (typically $N_l$=400 for the
Fermi gas and $N_l$=200 for the BEC). The system can be described
in terms of Bloch states of quasimomentum, whose energy spectrum
$E(q)$ is shown in Fig.~\ref{fig:band} for the first two bands for
a lattice with $s$=8. The width of the first band $\delta
E$=0.12$E_R$, is much smaller than the gap $\Delta E$=3.8$E_R$.
The ground state energy is half the oscillator quantum in each
lattice site $E_0$=$\hbar\omega_l/2$$\approx$2.49$E_R$, and
corresponds to rather large oscillating frequencies:
$\omega_l=2\pi\times 43000$~s$^{-1}$ and $\omega_l=2\pi\times
20000$~s$^{-1}$for K and Rb, respectively. Due to the presence of
the radial degrees of freedom, the system has actually bundles of
bands with identical dispersion, spaced by the radial energy
quantum $\hbar\omega_r\approx$0.038$E_R$ (dotted lines in
Fig.~\ref{fig:band}). We note that the degeneracy of the $n$-th
band is $n$+1, since two radial dimensions are involved.

While the condensate occupies macroscopically the state at $q$=0
in the fundamental radial band, identical fermions can have only
single occupancy of the $N_l$ momentum states in each band. At the
low temperatures of our system ($k_B T<\Delta E$) we can  restrict
our attention to the bundle of bands relative to the fundamental
band of the lattice. If we also neglect the band structure, i.e.
the motion along the lattice, since $\delta E\ll k_B T$, the
system has clearly a quasi-2D character, with the axial motion
within each lattice site frozen to the zero point motion. To
confirm this expectation, we have looked for a change of the
momentum distribution of the Fermi gas, since this is one of the
properties that can be more easily measured in the experiment. In
an harmonic potential of dimensionality $d$, with density of
states $g(E)\propto E^{d-1}$, the Fermi momentum at $T$=0 is
linked to the number of atoms by the relation $k_F\propto
N^{1/2d}$. In each quasi-2D lattice site we should therefore
expect a dependence $k_F\propto N^{1/4}$ of the radial Fermi
momentum in each lattice site, which is notably different from the
$k_F\propto N^{1/6}$ expected in 3D.

\begin{figure}
\centerline{\includegraphics[width=8.5cm,clip=]{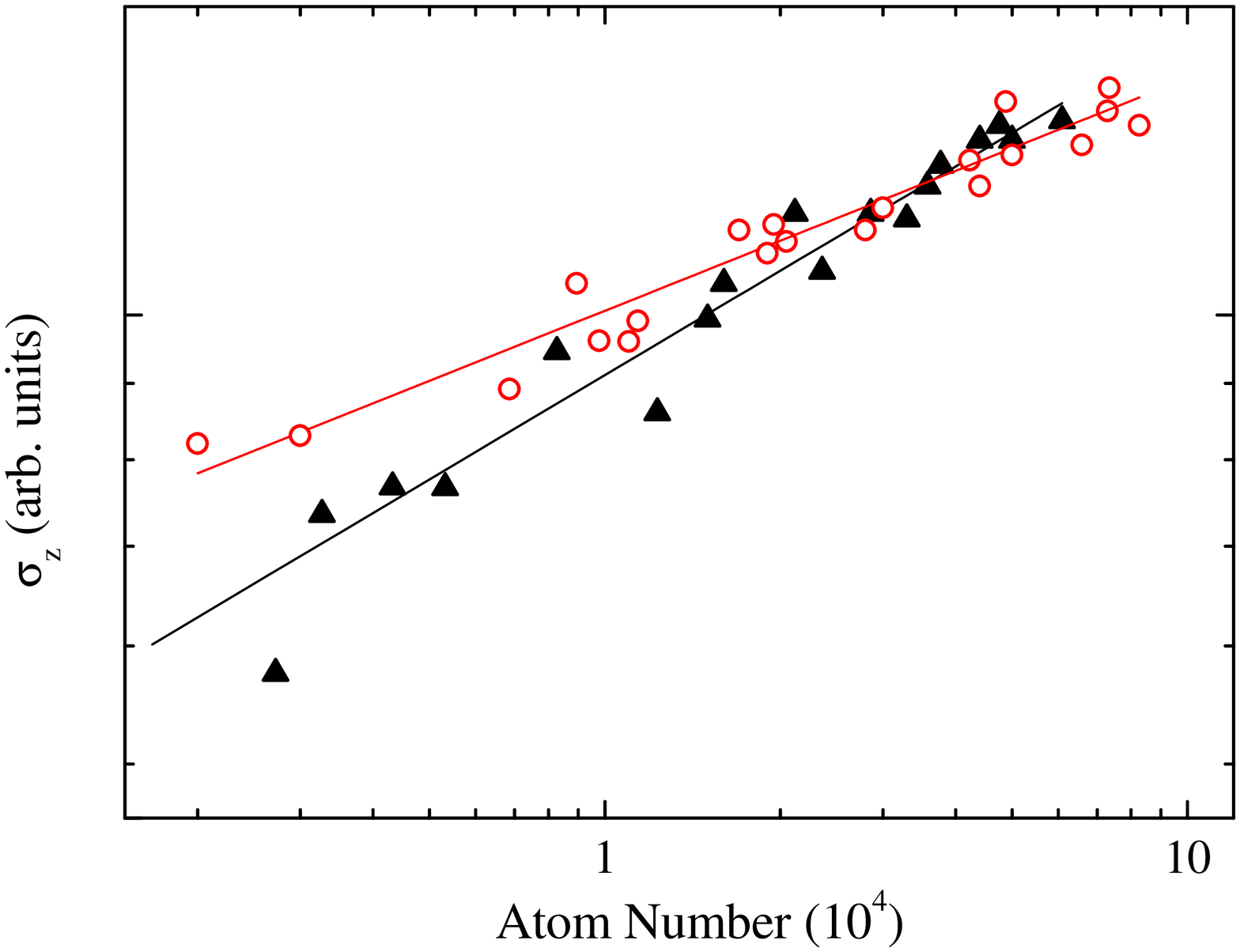}}
\caption{Width of the Fermi gas after ballistic expansion for the
magnetic potential (circles) and for the combined magnetic and
optical potential (triangles). The different power law in the
lattice is due to the reduced dimensionality of the Fermi gas.}
\label{fig:expansion}
\end{figure}

In the experiment we have measured the radial momentum
distribution of the Fermi gas in the lattice as a function of the
total number of atoms. This can be easily varied by adjusting the
initial loading of potassium in the magnetic trap. At an expansion
time $\tau$=8~ms$\gg\omega_r^{-1}$, the radial profile reflects
the initial momentum distribution. The experimental observation is
summarized in Fig.~\ref{fig:expansion}, where we compare the
radius of the Fermi gas released from the combined potential with
that released from the pure magnetic potential. Each radius is
measured with a gaussian fit to the central section (about 20
$\mu$m) of the cloud. The best fit of the two sets of data with
$\sigma_z(N)$=$aN^{1/n}$ curves gives $n$=4.3$\pm$0.3 and
$n$=6.1$\pm$0.6, respectively, confirming our expectation of
reduced dimensionality in the lattice \cite{noteenergy}.

The appropriateness of using this low-temperature model of the
system is confirmed by a one-dimensional Thomas-Fermi fit to the
radial profile of the clouds to determine the 2D temperature. For
the measurements of Fig.~\ref{fig:expansion} we find
$T^{2D}/T_F^{2D}\approx$0.4, where $k_B
T_F^{2D}$=$\hbar\omega_r(2N/N_l)^{1/2}$. We observe that the
reduced temperature does not change appreciably with $N$,
consistent with what we usually observe in the 3D trap. Note that
the minimum temperature is however higher than the one we usually
reach in 3D \cite{fermibose}.

We have repeated the same measurement with non-degenerate fermions
in the lattice at $T\approx$700~nK$>T_F$, where we have observed
no variation of the radius with $N$. This is consistent with the
expectation of a number-independent momentum distribution for a
classical gas, even in reduced dimensionality.

We can now take into account the contribution of the band
structure to the density of states. In the approximation of a
boxlike axial magnetic trapping, i.e. assuming that the fermions
are evenly distributed along the lattice, we have calculated the
dependence of the radial Fermi momentum on the atom number at
$T$=0. For high enough $N$ we can neglect the discreteness of the
radial energy levels, and we obtain a behavior $k_F\approx
N^{1/n}$, with $n$=4.2. This is very close to the result for 2D,
as expected because of the smallness of the axial energy $\delta
E\ll E_F$. Note that the atoms are loaded in the lattice at
$T\approx$0.5$T_F$, where also the 3D radius of the cloud begins
to be dependent on $N$. In the limiting case of loading at $T\ll
T_F$, the 3D Fermi radius scales as $N^{1/6}$, and computing the
number of atoms per site we would obtain an exponent $n$=4.8 for
the radial momentum.

An interesting question is how the sympathetic cooling varies
along the lattice. To check our ability to address small axial
sections of the Fermi gas, we have to consider the axial expansion
from the combined trap. The axial profile of the expanding cloud
is in general a convolution of the profile of each lattice site
and of the distribution along the sites. At expansion times
$\tau\gg\omega_l^{-1}$, i.e. when the individual site have merged,
the axial width of the cloud has the form $\sigma(\tau)\approx
R_F/2\sqrt{1+2(\sigma_l/R_F)^2\omega_l^2\tau^2}$, where
$R_F\propto N^{1/6}$ is the axial Fermi radius in the 3D trap and
$\sigma_l$ is the width of each site
$\sigma_l$=$\sqrt{\hbar/m\omega_l}\approx$76~nm. At the time
$\tau\approx$8~ms that we normally use in the experiment, the
width of each cloud initially confined in an individual site is
comparable to the overall width. Although the axial selectivity
would obviously be better at shorter times, here we estimate that
the lateral lattice sites give still only a small contribution to
the radial profile at the center of the cloud. For example, in the
case of a constant reduced temperature $T/T_F$ along the lattice,
the central section of the expanded cloud would be just 10\%
narrower than the individual central site. In the experiment we
typically measure a constant radius along the lattice, which
instead indicates a decreasing degeneracy for lattice sites far
from the center. A reduction of the efficiency of sympathetic
cooling in a tight lattice is actually not unexpected, since the
bosons in the lateral lattice sites are rapidly removed by the
evaporation, because of their larger Zeeman energy, and they are
not very efficiently replaced through tunnelling. As soon as their
number drops below that of the fermions, the cooling slows down,
and eventually stops when all the bosons are removed from the
site.

One of the issues of the experiment was to understand whether a
tight lattice would affect the stability of the degenerate
mixture, since in principle collapse or 3-body decay
\cite{collapse} can be favored by an increased confinement
\cite{roth}. However, we were not able to observe a collapse of
the mixture in the combined potential for $s$=8, even for total
atom numbers approaching the critical values in 3D. Moreover, in
the lattice we measured a lifetime of the Fermi gas
$\tau\gtrsim$0.5~s, limited by 3-body collisions with Rb atoms,
which is comparable to the one in the magnetic potential
\cite{collapse}. Both these observations seem to indicate that the
effective density overlap of the BEC and Fermi gas in the lattice
is not increased with respect to that in the magnetic potential
alone. We have compared these observations with the prediction of
our simple model which neglects both the axial confinement and the
axial motion of the atoms along the lattice, i.e. assumes both
bosons and fermions confined to quasi-2D pancakes. The density
distribution of both species has therefore a 2D Thomas-Fermi
radial shape appropriate to the statistics \cite{noteradii}, and a
gaussian axial shape, with rms width $\sigma_l$. For the typical
atom numbers, we find that the peak density of the Fermi gas
increases by approximately 5 with respect to the pure magnetic
trap, while the increase of the condensate peak density is limited
to approximately 1.5 by the boson-boson mean-field repulsion. The
overlap of the two species in the lattice is however worse, mainly
because the radius of the Fermi gas decreases by almost 50\%
passing from 3D to 2D, thus increasing the effect of the
differential gravitational sag \cite{fermibose}. As an example,
the radii of a BEC and a Fermi gas containing 5$\times$10$^4$
atoms each are 2.8~$\mu$m and 4~$\mu$m, respectively, hence
comparable to the gravitational sag $\delta z$=3.6~$\mu$m. These
predictions confirm qualitatively both observations on the
stability of the mixture and on the slightly reduced efficiency of
the sympathetic cooling.

\begin{figure}
\centerline{\includegraphics[width=8.5cm,clip=]{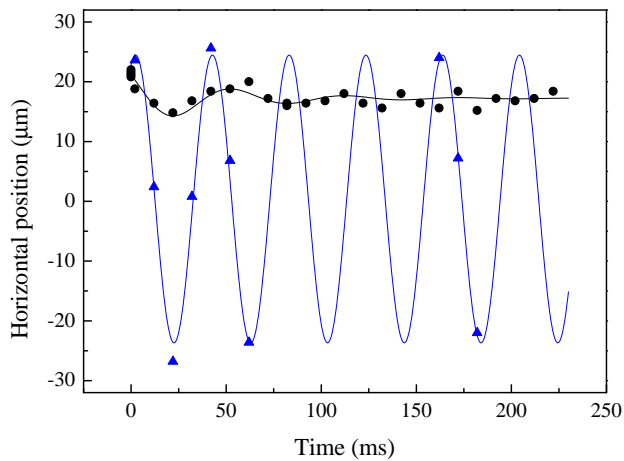}}

\caption{Dipolar axial oscillations of a Fermi gas in the combined
magnetic potential and optical lattice with $s$=7.6 (circles), and
in the magnetic potential alone (triangles). The oscillations are
excited by displacing the magnetic trap by 15~$\mu$m, and the
position of the cloud is detected after 8~ms of ballistic
expansion. In presence of the lattice the Fermi gas performs a
strongly damped oscillation close to the equilibrium position in
the magnetic trap prior to the displacement.} \label{fig:sloshing}
\end{figure}

The different statistical nature of bosons and fermions determines
also their \textit{dynamics} in the lattice. We have studied the
dipolar oscillations of either the Fermi gas or the BEC in the
combined potential \cite{evap}, which are induced by a sudden
axial displacement of the magnetic trap. As shown in
Fig.~\ref{fig:sloshing}, the oscillation of the Fermi gas is
strongly damped in presence of the tight lattice, and the center
of mass of the cloud is hardly displaced from the original
equilibrium position. Differently, the BEC oscillation in the
lattice proceeds without damping, as originally studied in
Ref.~\cite {cata}.

To understand these different dynamical behaviors we can no longer
neglect the axial magnetic potential. In a semiclassical picture,
the atoms move under the force exerted by the harmonic magnetic
potential, with the Bloch velocity dispersion $v(q)$=$1/\hbar\;
\partial E/\partial q$ set by the lattice. The narrow quasimomentum
distribution of the BEC allows an undamped dipolar oscillation,
provided that $q$ is confined to the parabolic part of the
fundamental energy band in Fig.~\ref{fig:band}, which has an
overall cosine shape in this tight-binding regime. This picture
changes completely for the Fermi gas, because of its broad
quasimomentum distribution. Already at equilibrium, i.e. before
exciting the dipolar motion, most of the individual fermions
oscillate along a wide section of the fundamental bands during
their motion in the trap, and therefore in general they perform
strongly non-harmonic oscillations. This single-particle behavior
will clearly affect also the collective dipolar oscillation, which
is likely to be damped, as it would happen in an anharmonic
potential. If the atoms are no more able to perform full
oscillations in the magnetic potential, one could also expect that
the Fermi gas is eventually blocked to an equilibrium position far
from the magnetic trap minimum, as we observe in the experiment.

To access a regime where the Fermi gas can perform dipolar
oscillations with larger amplitude, one should reduce either the
number of atoms, in order to have a much narrower momentum
distribution, or the lattice depth, in order to have almost
parabolic bands. In the experiment we have actually observed that
the oscillation amplitude of the Fermi gas grows as the lattice
height is lowered.  We plan to further investigate the
oscillations of the Fermi gas in shallow lattices, and to make a
comparison to those of a thermal cloud of bosons, to study
possible effects of the statistics.

We have also studied the oscillation of the mixture of BEC and
Fermi gas in the tight lattice, seeking possible interaction
effects. On the BEC side, we have observed just a moderate damping
due to collisions with fermions, as we already measured in the
magnetic trap \cite{sloshing}, and no noticeable frequency shift.
In contrast, in presence of the BEC, the oscillation of the Fermi
gas in the lattice appears to be further reduced, and becomes no
longer detectable. Future work on shallower lattices will also
help to elucidate the effects of the interactions with bosons on
the oscillations of the Fermi gas.

In conclusion, we have produced a Fermi gas of atoms in a 1D
lattice by sympathetic cooling with bosonic atoms. The realization
of quasi-2D Fermi gases in the lattice is interesting in view of
achieving fermionic superfluidity, as discussed in Ref.
\cite{petrov}. Our results are promising also for loading the
Fermi gas or the Bose-Fermi mixture in 2D or 3D lattices. These
are of great interest for studies on high-$T_c$ superfluidity
\cite{cirac}, novel quantum phases \cite{albus}, and the formation
of ultracold heteronuclear molecules \cite{damski}.

We acknowledge contributions by E. de Mirandes and useful
discussions with the BEC theory group in Trento. This work was
supported by MIUR, by EC under contract HPRICT1999-00111, and by
INFM, PRA ``Photonmatter'' R.H. was supported by EC with the
program Marie Curie Training Sites, under contract HPMT2000-00123.


\begin{thebibliography}{99}
\bibitem{noterecoil} $E_R/k_B$ is 415~nK and 190~nK for K and Rb,
respectively.
\bibitem{noteenergy} Note that
even at the lowest number of atoms the 3D Fermi energy is expected
to be larger than $\delta E$: $E_F\approx 0.25E_R$ for
$N\approx$10$^3$.
\bibitem{evap} The bosons can be completely removed from the trap
with a RF sweep at the end of the evaporation.
\end{thebibliography}
\end{document}